\documentclass[10pt,letterpaper,twocolumn]{article}

\usepackage{ol2}
\usepackage[draft]{hyperref}

\usepackage{amsmath}
\usepackage{epsfig}
\usepackage{amssymb}
\usepackage{color}
\usepackage{graphicx}

\newcommand{\eqb}{\begin{eqnarray}}
\newcommand{\eqe}{\end{eqnarray}}

\newcommand{\tkap}{\tilde{\kappa}}

\newcommand{\tg}{\tilde{g}}

\newcommand{\PT}{{\cal PT}}

\def\bes{\begin{subequations}}
\def\ees{\end{subequations}}

\begin{document}

\twocolumn[

\title{Conservative and $\PT$-symmetric compactons in waveguide networks.}

\author{A. V. Yulin and  V. V. Konotop}
\address{
Centro de F\'isica Te\'orica e Computacional and Departamento de F\'isica, Faculdade de Ci\^encias, Universidade de Lisboa, Avenida Professor Gama Pinto 2, Lisboa 1649-003, Portugal
}
\begin{abstract}
Stable discrete compactons in arrays of inter-connected three-line waveguide arrays are found in linear and nonlinear limits in conservative and in parity-time ($\PT$) symmetric models. The compactons result from the interference of the fields in the two lines of waveguides ensuring that the third (middle) line caries no energy. $\PT$-symmetric compactons require not only the presence of gain and losses in the two lines of the waveguides but also complex coupling, i.e. gain and losses in the coupling between the lines carrying the energy and the third line with zero field. The obtained compactons can be stable and their branches can cross the branches of the dissipative solitons. Unusual bifurcations of branches of solitons from linear compactons are described.

\end{abstract}



]

Optical waveguide arrays are the systems allowing for efficient control of light propagation~\cite{review}. Such devices can work in both linear and nonlinear regimes, the latter one being of particular relevance for situations where the energy flux has to be concentrated in a finite spatial domain. The role of the nonlinearity consists in counter-balancing diffraction which is an intrinsic property of generic discrete systems. In the same time the nonlinearity results in the interaction between the modes thus limiting possibilities of their practical usage.

Thinking about the energy localization in an array, one naturally should explore the possibilities of the existence of the {\em compacton solutions}, as excitations whose {\em field is concentrated in a given finite domain and is exactly zero outside this domain}. Compactons were first predicted for continuous systems~\cite{compact_cont_1,compact_cont_2} and later  found~\cite{KonTak} and explored (see e.g.~\cite{compact_disc_1,compact_disc_2,compact_disc_3,compact_disc_3_1,compact_disc_4}) in nonlinear lattices. The genuine bright compactons reported so far had essentially nonlinear nature. Due to the mentioned diffractive properties of lattices observation of compactons in the linear regime is usually considered to be impossible.
In the present Letter we show
that this is not necessarily so, and  it is possible to construct simple networks where {\em compactons can exist in pure linear regime}, i.e. they appear  as eigenmodes of the system, and thus can propagate along the structure without distortion.
Moreover, the array can be designed in such a way that the compacton modes are not affected by the nonlinearity, extending the results to arrays of Kerr-nonlinear waveguides.   Furthermore, we show that the system can be enriched by considering active and lossy waveguides, still preserving compacton solutions. In this last case the consideration will be limited here to the particular case  where the gain and losses are balanced, ensuring parity-time ($\PT$) symmetry~\cite{Muga,Ruter}, giving origin to $\PT$-compactons.

\begin{figure}
\includegraphics[width=0.9\columnwidth]{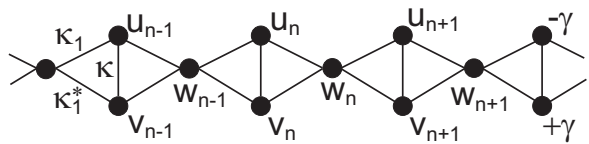}%
 \caption{Schematic representation of the array with the notations used in the text. Black dots and line designate the waveguides and links among them.}%
 \label{fig1}
\end{figure}

We consider an array of coupled optical waveguides whose geometry is symmetric with respect to the central-line [Fig.~\ref{fig1}]. We believe that these geometrically complicated structures can be manufactured either by writing the waveguides in silicon matrix using femtosecond lasers \cite{heinrich} or by other technologies, see \cite{tanya} and \cite{silicon}.

The equations governing the beam propagation
for the dimensionless electric fields in the upper--  ($u_n$), lower-- ($v_n$), and middle--  ($w_n$) line waveguides read
\begin{subequations}
\label{main-conservative}
\begin{eqnarray}
\label{eq_u}
i\dot{u}_n=\kappa v_n-i\gamma u_n+\kappa_1(w_{n-1}+w_{n})+ g|u_n|^2u_n
\\
\label{eq_v}
i\dot{v}_n=\kappa u_n+i\gamma v_n+\kappa_1^*(w_{n-1}+w_{n})+ g|v_n|^2v_n
\\
i\dot{w}_n=
\kappa_1 (u_n+u_{n+1})+\kappa_1^*(v_{n}+v_{n+1})+\tg |w_n|^2 w_n
\end{eqnarray}
\end{subequations}
Here an overdot stands for the derivative with respect to evolution coordinate $z$, asterisk means complex conjugation, $\kappa$ (it is considered positive) and $\kappa_1=\tkap  e^{i\phi/2}$ with real $\tkap$ and $\phi$, are the coupling constants (that can be complex, see \cite{flach}) between the respective pairs of arrays;
$g$ and $\tg$  are the Kerr nonlinearities of the $u$ and $v$ waveguides and of the $w$ waveguides, respectively.
In a general situation $u$ and $v$ waveguides will be considered absorbing and active with balanced gain and loss characterized by the coefficient $\gamma$, i.e. having parity-time ($\PT$) symmetric structure. Notice that at $\kappa_1=0$ the system is reduced to a sequence of decoupled nonlinear $\PT$-symmetric dimers which have been well studied in the linear~\cite{Kulishov,Guo,Ruter} and nonlinear~\cite{Ramezani,Kevrekid} settings.

We start with the dispersion relation of the linear waves, i.e. use the anstaz $(u_n,v_n, w_n)=(u,v,w) e^{i(bz-kn)}$ in (\ref{main-conservative}) at $g=\tg=0$. This yields the equation which   is convenient to write down in the form
\begin{eqnarray}
\label{dispers}
b^3-(\kappa^2-\gamma^2)b
-8\tkap^2 [b\cos \phi+\gamma\sin\phi+\kappa] \cos^2 \frac{k}{2} =0
\end{eqnarray}
A peculiarity of this dispersion relation is that it allows for $k$ independent solutions for the propagation constant $b=b_0$ (i.e. having $db_0/dk=0$), i.e. for the diffractionless modes. Theses are modes
for which $b^3-(\kappa^2-\gamma^2)b=0$ and $b\cos \phi+\gamma\sin\phi+\kappa=0$ are satisfied simultaneously what occurs in the following two cases:
\begin{eqnarray}
\label{case1}
b_0=0 \quad\mbox{if}\quad \gamma=-\kappa/\sin\phi
\\
\label{case2}
b_0=-\kappa\cos\phi \quad\mbox{if}\quad \gamma=-\kappa\sin\phi
\end{eqnarray}
In the first case ($b_0=0$) the filed in the central waveguide can be arbitrary and it defines the fields in lossy and active waveguides:
\begin{subequations}
\label{case1_mode}
\begin{eqnarray}
u=\frac{\tkap\sin\phi}{\kappa\cos^2\phi}\left(1+e^{ik}\right)\left(ie^{i\phi/2}+  e^{-i\phi/2}\sin\phi\right)w
\\
v=\frac{\tkap\sin\phi}{\kappa\cos^2\phi}\left(1+e^{ik}\right)\left(ie^{-i\phi/2}+ e^{i\phi/2}\sin\phi\right)w
\end{eqnarray}
\end{subequations}

In the  case (\ref{case2}) we obtain a {\em dipole mode} in which the fields in the $u-$ and $v-$ waveguides are given by
\begin{eqnarray}
\label{case2_mode}
 u_{dip}=\alpha v_{dip}, \quad \alpha=\frac{\kappa^2-\tkap^2(1+i\cos k)}{ \tkap^2(1+i\cos k)-\kappa^2\cos\phi}e^{i\phi}
\end{eqnarray}
while $w-$ waveguides carry no energy ($w=0$) due to the destructive interference of the  $u$-- and $v$-- fields.
From the last formula we observe that the dipole mode has a simple conservative limit  $\gamma=\kappa=\phi=0$, for which which $\alpha=-1$, i.e. $u_{dip}=-v_{dip}$ [such a limit obviously does not exist for the mode (\ref{case1_mode})]. Since  $k$ does not enter in the solution (\ref{case2_mode}) we conclude that even in the linear case there exists a {\em compacton} solution in which only one (or a finite number) of waveguides carry the field.

\begin{figure}
\epsfig{file=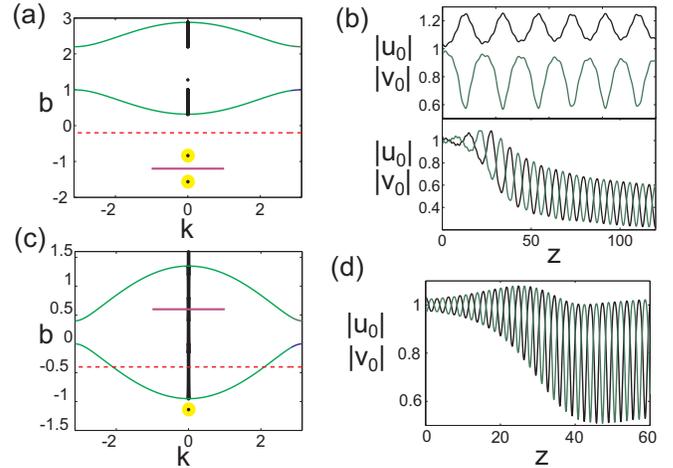,angle=0,width=\columnwidth}
\caption{(a) The dispersion characteristics for $\kappa_1=0.4$, $\kappa=1.2$. The horizontal dashed line indicates $b_0$, while the solid green lines  show the dispersion curves $b_\pm$.  Thick short magenta line marks the position of the dimer frequency. The black dots show the frequencies of the linear excitations in the presence of an excited dipole mode.   (b) The dynamics of the unstable modes for $\kappa=0.6$ (upper panel) and $\kappa_1=1.0$ (lower panel). (c) The same as panel (a) but for $\kappa_1=0.4$, $\kappa=0.4$. (d) The dynamics of the unstable mode for $\kappa_1=0.6$.
}
\label{fig2}
\end{figure}

Let us now study the conservative compactons in more details letting $\kappa_1$ to be real and $\gamma=0$ (i.e. letting $\phi=0$). Then (\ref{dispers}) can be solved explicitly:
\begin{eqnarray}
\label{eq:disp_rel}
b_0=-\kappa, \quad b_\pm=\frac{\kappa}{2}\pm\frac 12 \sqrt{\kappa^2+16\kappa_1^2(1+\cos k)}
\end{eqnarray}
These branches of the spectrum are illustrated in Fig.~\ref{fig2}.  According to (\ref{case2_mode}) in the linear limit ($g,\tg=0$) for the dipole mode $w_n=0$  and Eqs.~(\ref{eq_u}), (\ref{eq_v}) are decoupled, the solution can be written down in the form $(u_n,v_n,w_n) =c_n(1,-1,0)$, where $c_n$ ($n$ is integer) is an arbitrary set of numbers.
One can further particularize the choice letting $c_0\neq 0$ and $c_n=0$ for all $n\neq 0$. This gives an {\em exact compacton} solution in which the light propagates along the "effective dimer" of the $(u_0,v_0)$ waveguides. Clearly, in the infinite network there can exist an infinite number of compactons each one having an independent amplitude but the same propagation constant given by (\ref{case2}).

The {\em symmetric} modes corresponding to the branches $b_\pm$ are given by:
\begin{equation}
\label{mode_pm}
u_\pm=v_\pm=\alpha_\pm w, \quad \alpha_\pm=\frac{\sqrt{\kappa^2+16\kappa_1^2(1+\cos k)}\pm\kappa}{4\tkap \left(1+e^{-ik}\right)}
\end{equation}
where $w$ is chosen arbitrarily.

A remarkable fact is that the conservative compactons persist in the nonlinear case. Indeed one ensures that the dipole-mode compacton $(u_n^{(c)},v_n^{(c)},w_n^{(c)})=(1,-1,0)\delta_{n0} \rho e^{ib z}$ where $b=-\kappa-g\rho^2$ and $\delta_{n,0}$ is the Kronecker delta, is a solution of (\ref{main-conservative}) at $\gamma=\phi=0$.

 To address the stability of this nonlinear compacton we use the ansatz $u_n=u_n^{(c)}+\tilde u (z)$,  and $v_n=v_n^{(c)}+\tilde v_n(z)$ and linearize the system (\ref{main-conservative}) with respect to small  $\tilde u_n$, $\tilde v_n$. It turns out, however, that the analysis is more conveniently performed in the variables $\xi_n=\tilde u_n + \tilde v_n$ and $\eta_n=\tilde u_n - \tilde v_n$ satisfying
\begin{eqnarray}
\label{eq:eta_lin_stab}
i\dot{\eta}_0=g\rho^2 (\eta_0+\eta_0^*),
\\
 \label{eq:xi_lin_stab1}
 i\dot{\xi}_n=2\kappa\xi_n+\delta_{n0} g\rho^2( \xi_n+ \xi_n^*)
 +2\kappa_1(w_{n-1}+w_n),
   \\
 \label{eq:xi_lin_stab2}
i\dot{w}_n=\kappa_1(\xi_n+\xi_{n+1})
\end{eqnarray}
(notice that $\eta_n=0$ for $n\neq 0$).
The equation for $\eta_0$ is decoupled giving gets the zero eigenvalue (it can be associated with the phase symmetry of the system). Thus, the instability of the system is determined by Eqs.~(\ref{eq:xi_lin_stab1}), (\ref{eq:xi_lin_stab2}). In the limiting case $\kappa_1=0$ the coupling also disappears from these equations and using the anstaz  $\xi_0 =\tilde \xi_0 e^{i \Delta z} + \bar{\xi}_0 e^{-i\Delta z}$ one  obtains from (\ref{eq:xi_lin_stab1}):
$
\Delta^2=8\kappa(2\kappa+g\rho^2).
$
Thus in the absence of the coupling the dipole mode is stable for $g>0$ (recall that $\kappa>0$). If $g<0$ then the mode is stable under the condition  $|g|\rho^2<\kappa$ meaning that the absolute value of the propagation constant is less than $b_{cr}=4\kappa$.

For $\kappa_1\neq 0$ we solve numerically the eigenvalue problem  (\ref{eq:xi_lin_stab1}), (\ref{eq:xi_lin_stab2}). We start the discussion with the case $g=-1$ when the dipole mode is stable at $\kappa_1=0$. Fig.~\ref{fig2}(a) shows the dispersion characteristics with overlapped spectrum of the linear excitations in the presence of the dipole mode for relatively small $\kappa_1$.
At some value of $\kappa_1$ two real eigenvalues (marked by yellow circles in (a)) collide and produce an instability. The critical value of $\kappa_1$ can be calculated analytically because two eigenvalues belonging to the discrete spectrum are involved. If we increase the coupling $\kappa_1$ even further then at some point the central gap closes and a pair of eigenvalues generating the instability transforms into a quartet. Fig.~\ref{fig2} (b) shows the development of instability for $\kappa_1=0.6$ and $\kappa_1=1.0$. Remarkably, the compacton is not destroyed completely but evolves towards an oscillating state.

We conclude the analysis of conservative compactons by considering the case $g>0$. Now the dipole mode is unconditionally stable for zero coupling  $\kappa_1=0$. However at some  $\kappa_1$ the discrete eigenvalues hit the continuum generating a quartet of eigenvalues and destabilizing the compacton. This case is illustrated in panels (c) and (d).

\begin{figure}
\epsfig{file=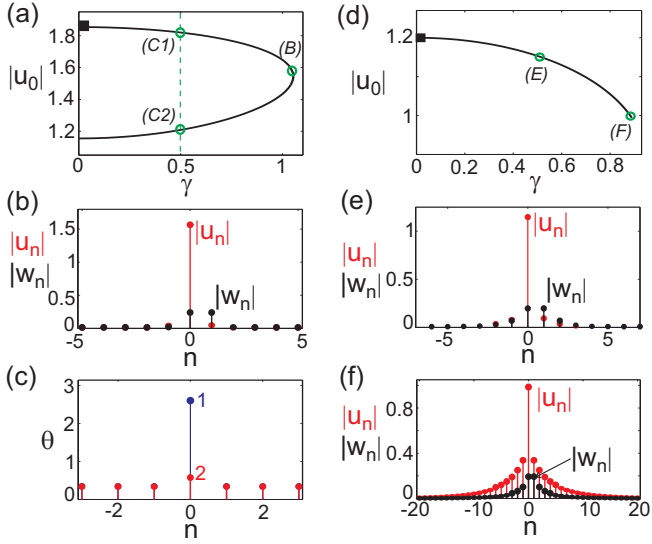,angle=0,width=\columnwidth}
\caption{The dependencies of the maximum absolute value of the field $u_n$ on the dissipation $\gamma$ for the solitons bifurcating from the compacton at $\gamma=0$ are shown in panels (a) and (d) for structures with the propagation constants $b_s=0.44$ and $b_s=2.44$ correspondingly. The bifurcation points are marked by the squares. Panel (b) shows the distribution of the fields $u_n$ (red circles) and $w_n$ (black circles) for the point marked on the bifurcation curve on panel (a) as { \it (B)}. Panel (c) shows the distribution of the mutual phase of the fields $u_n$  and $v_n$ defined as $\theta=arg(v_n u_n^{*})$. The distributions marked as $1$ and $2$ correspond to the points {\it C1} and {\it C2} in panel (a).  Panels (e) and (f) show the distributions of the fields $u_n$ and $w_n$ in the solitons corresponding to the points {\it (E)} and {\it (F)} on the bifurcation curve shown in panel (d). The parameters are  $g=1$, $\kappa=1$, $\kappa_1=0.25$. }
\label{fig3}
\end{figure}

Now we turn to the analysis of the $\PT$-symmetric case. First of all we notice that in the case of pure real coupling $\kappa_1$ ($\phi=0$) the dispersion relation (\ref{dispers}) still may have three real roots for any $k$. This corresponds to the unbroken $\PT$-symmetric phase, which exists in the following parameter range:
\begin{eqnarray}
\label{cond:disp_real_1}
 2\kappa_1^2<\kappa^2,
\qquad
\gamma^2<\kappa^2+4\kappa_1^2-3\left( 2 \kappa \kappa_1^2 \right) ^{2/3}.
\end{eqnarray}
However for $\gamma \neq 0$ the respective dispersion characteristics  can never have a nondiffractive branch with the curvature equal to zero for all $\kappa$. In other words no compactons are possible for nonzero losses. Instead one can observe bifurcation of the solitons from the linear modes. Interestingly, these dissipative soliton branches bifurcate from the compactons when nonzero $\PT$-symmetric terms appear.
The bifurcation diagram for the soliton branches for fixed propagation constants $b_s$ are
is shown in Fig.~\ref{fig3} (a).  The point where solitons bifurcate from the compacton (antisymmetric dipole mode) is shown by the black square on the vertical axis. With the increase of $\gamma$ the localized solution becomes wider, see [panel (b)] where the distributions of $u_n$ and $w_n$ are shown for the point {\it (B)} on the bifurcation curve in Fig.~\ref{fig3}~(a).

At a certain point the soliton looses stability, then the bifurcation curve has a fold and turns back. Finally the bifurcation curve reaches the vertical axis which corresponds to coming back to the conservative case. However the second conservative state corresponds to a symmetric state and so it is not a compacton but a soliton (as it is discussed above the compact solution is possible for the antisymmetric dipole modes only).

The difference between the upper and the lower branches of the bifurcation curve is illustrated in panel (c) showing the relative phase between the fields $u_n$ and $v_n$ at the points { \it (C1)} and { \it (C2)}. One can see that for { \it (C1)} the phase $\theta$ is relatively close to zero implying that the pairs $(u_n,v_n)$ are in a quasi-symmetric state. On contrary, for { \it (C2)} the phase is closer to $\pi$ which means that $u_n$ and $v_n$ are in a quasi-antisymmetric state. For $\gamma=0$ we have either pure symmetric or antisymmetric state.

If the intensity of  dissipative solitons belonging to a given branch goes beyond some threshold value, as $\gamma$ grows, then the continues transformation from antisymmetric conservative compacton into symmetric soliton becomes impossible. The corresponding bifurcation diagram is shown in Fig.~\ref{fig3}~(d) and the absolute values of the fields are illustrated in panels (e) and (f). One can see that the soliton is getting wider and the numerical simulations signal that at the end of the bifurcation diagram the soliton becomes infinitely wide. This singularity is responsible for the break of the bifurcation diagram.

Remarkably if the coupling between the sites is not pure conservative then compact solutions are possible in the $\PT$-symmetric case as well. To have the compact solution one has to ensure that there can propagate a non-diffractive beam, i.e. that $b$ is independent on $k$.
Recalling that in the case (\ref{case2}) the diffractionless mode has $w=0$ we obtain that  the excited $\PT$-symmetric dipole mode has the same structure as the linear $\PT$-compacton: the nonlinearity changes the frequency but leaves untouched the relation between the fields $u$ and $v$.

\begin{figure}
\epsfig{file=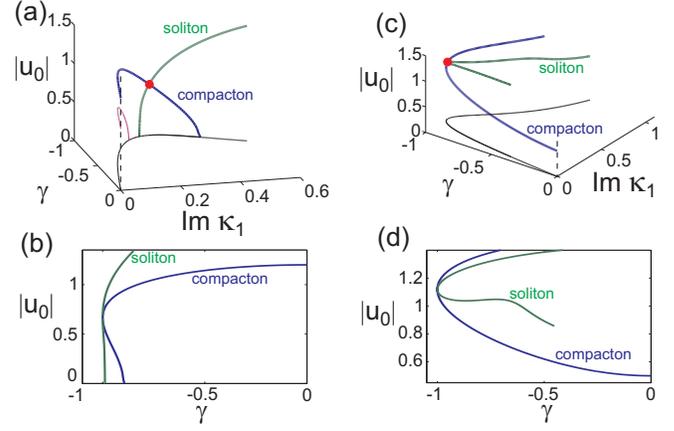,angle=0,width=\columnwidth}
\caption{The thinner black curve in panel (a) shows the trajectory in $\gamma$-$Im \kappa_1$ plane along which the compact solution can exist. The thicker blue ($b_s=0.44$) and thinner magenta lines ($b_s=-0.36$) show the bifurcation diagrams of the compacton. The bifurcation diagram of the soliton is shown by the green line. In panel (b) the bifurcation diagrams of the compactons and the soliton are shown as  functions of $\gamma$. Panels (c) and (d) show the same but for the negative nonlinearity $g=-1$ and the propagation constant $b_s=-1.25$. The other parameters are  $\kappa=1$, $\kappa_1=0.25$. }
\label{fig4}
\end{figure}

Let us now study the bifurcation of these compactons for $g=1$. To have a compact solution we have to follow along the line $\gamma=-k\sin\phi$ in the parameter plane as shown in    Fig.~\ref{fig4} (a). Departing from the conservative case we follow the line tracing the maximum amplitude of $u-$field (thicker blue line). The projection of the bifurcation curve on $(|u_0|,\gamma)$ plane is shown in Fig.~\ref{fig4} (b). The solution remains compact with only one "cell" $(u_0,v_0)$ excited. Since the compact solution do not interact to each other one can use the dipole modes as building blocks for designing infinitely many compact, periodic or non-periodic, solutions.

Going along the bifurcation curve we arrive to the bifurcation point where the compact and soliton solutions collide, this point is indicated by the red circle in panel (a). The bifurcation curve of the soliton along the curve of the existence of the compactons is shown by the green line. The same curve in $(|u_0|,\gamma)$ plane is shown in panel (b). It is worth noticing that the compacton and the soliton solutions can coexist. The same happens for the case of negative nonlinearity $g=-1$, see panels (c) and (d).

To conclude we briefly summarize the main results reported in the paper.  It is demonstrated that compactons in arrays of connected waveguides can exist in linear and nonlinear limits, in conservative and $\PT$-symmetric models. The compacton solutions result from the interference of the fields in two lines of the waveguides ensuring zero energy carried in the third-line (the dipole modes). $\PT$-symmetric compactons require not only the presence of gain and losses in waveguide lines  but also complex coupling coefficients, i.e. gain and losses in the coupling between the lines carrying the energy and the the third line of waveguides.The obtained $\PT$-symmetric compactons can be stable and their branches can cross the branches of the dissipative solitons. Being stable and strongly localized objects the reported discrete compactons can be of particular interest for such applications as optical memory, optical switch, phase synchronization, and others. Finally we remark that the considered discrete compactons can also be generalized for two dimensional and for nonlinear dissipative (not {\it PT}-symmetric) cases that will be reported elsewhere.

The work was supported by the  FCT (Portugal) grants: PTDC/FIS/112624/2009,    PEst-OE/FIS/UI0618/2011 and PTDC/FIS-OPT/1918/2012.


\begin{thebibliography}{99}

\bibitem{review} F. Lederer, G.I. Stegeman, D.N. Christodoulides, G. Assanto, M. Segev, and Y. Silberberg,
Phys. Rep. {\bf 463}, 1 (2008)

\bibitem{compact_cont_1} P. Rosenau and J.M. Hyman, Phys. Rev. Lett. {\bf 70}, 564 (1993).

\bibitem{compact_cont_2} P. Rosenau, Phys. Rev. Lett. {\bf 73}, 1737 (1994);


\bibitem{KonTak} V.V. Konotop and S. Takeno,
Phys. Rev. E, {\bf 60}, 1001 (1999).

\bibitem{compact_disc_1} E. Coquet, M. Remoissenet, and P.T. Dinda, Phys. Rev. E, {\bf 62}, 5767 (2000);

\bibitem{compact_disc_2}
M. Eleftheriou, B. Dey, and G.P. Tsironis, Phys. Rev. E, {\bf 62}, 7540 (2000);

\bibitem{compact_disc_3}
P.G. Kevrekidis and V.V. Konotop,    Phys. Rev. E, {\bf 65} 066614 (2002).

\bibitem{compact_disc_3_1} P.G. Kevrekidis, V.V. Konotop, and S. Takeno,  Phys. Lett. A, {\bf 299}, 166  (2002).

\bibitem{compact_disc_4}
F.Kh. Abdullaev, P.G. Kevrekidis, and M. Salerno,
Phys. Rev. Lett.  {\bf 105}, 113901 (2010).

\bibitem{Muga} A. Ruschhaupt, F. Delgado, and J. G. Muga,
J. Phys. A  {\bf 38}, L171 (2005).

\bibitem{Ruter} C. E. R\"uter, K. G. Makris, R. El-Ganainy, D. N. Christodoulides,  M. Segev, and D. Kip,
Nature Phys.  {\bf 6}, 192  (2010)

\bibitem{heinrich} M. Heinrich, R. Keil, F. Dreisow, A. Tunnermann, A. Szameit, S. Nolte, Appl Phys B 104:469–480, (2011)

\bibitem{tanya} T.M. Monro and H. Ebendorff-Heidepriem, Annu. Rev. Mater. Res., 36:467–95, (2006)

\bibitem{silicon} P. Koonath and B. Jalali, Optics Express, Vol. 15, No. 20,  12686, (2007)

\bibitem{flach} N. V. Alexeeva, I. V. Barashenkov, K. Rayanov and S. Flach,  http://xxx.tau.ac.il/abs/1308.5862

\bibitem{Kulishov} M. Kulishov, J. Laniel, N. B\'elanger, J. Aza\~na, and D. Plant,  
Opt. Expr. {\bf 13}, 3068 (2005)

\bibitem{Guo} R. El-Ganainy, K. G. Makris,  D. N. Christodoulides, and Z. H. Musslimani,    
Opt. Lett., {\bf 32}, 2632 (2007).



\bibitem{Ramezani} H. Ramezani, T. Kottos, R. El-Ganainy, and D. Christodoulides,
Phys. Rev. A, {\bf 82}  043803 (2010)

\bibitem{Kevrekid}   P.G. Kevrekidis, D.E. Pelinovsky, and D.Y. Tyugin,
SIAM J. Appl. Dynam. Syst. {\bf 12} (2013), 1210--1236.

\end{thebibliography}
\end{document}